# Estimation of electric field impact in deep brain stimulation from axon diameter distribution in the human brain

Johannes D. Johansson

Department of Biomedical Engineering, Linköping University, 581 85 Linköping, Sweden

johannes.johansson@liu.se

johannesj@gmail.com

## Abstract

Background: Finite element method (FEM) simulations of the electric field magnitude (EF) are commonly used to estimate the affected tissue surrounding the active contact of deep brain stimulation (DBS) leads. Previous studies have found that DBS starts to noticeably activate axons at approximately 0.2 V/mm, corresponding to activation of 3.4 µm axons in simulations of individual axon triggering. Most axons in the brain are considerably smaller however, and the effect of the electric field is thus expected to be stronger with increasing EF as more and more axons become activated.

Objective: To estimate the fraction of activated axons as a function of electric field magnitude.

Methods: The EF thresholds required for axon stimulation of myelinated axon diameters between 1 and 5 µm were obtained from a combined cable and Hodgkin-Huxley model in a FEM-simulated electric field from a Medtronic 3389 lead. These thresholds were compared with the average axon diameter distribution from literature from several structures in the human brain to obtain an estimate of the fraction of axons activated at EF levels between 0.1 and 1.8 V/mm.

Results: The effect of DBS is estimated to be 41·$EF$ – 7.4 % starting at a threshold level $EF_{t0}$ = 0.18 V/mm.

Conclusion: The fraction of activated axons from DBS in a voxel is estimated to increase linearly with EF above the threshold level of 0.18 V/mm. This means linear regression between EF above 0.18 V/mm and clinical outcome is a suitable statistical method when doing improvement maps for DBS.

## Abbreviations

DBS    Deep Brain Stimulation

EF    Electric Field magnitude

FEM    Finite Element Method

VTA    Volume of Tissue Activated

GPi    Globus Pallidus interna;

STN    Subthalamic Nucleus;

## Introduction

Deep brain stimulation (DBS) is a well-established technique for symptomatic relief in movement disorders such as Parkinson's disease, essential tremor, and cervical dystonia [1-3]. The exact mechanism of DBS is not fully known. However, there is experimental evidence that axons are triggered by the DBS pulses and are able to keep up with a pulse frequency of 120 Hz but the synapses at the end of them start to become depleted at 60 Hz. This depletion occurs at a faster rate for higher pulse frequency [4, 5]. DBS thus appears to act as a jammer of neural activity, which could explain why the clinical effect of DBS is similar to tissue-destroying lesioning for the commonly treated symptoms [6-9].

By comparing results from experimental studies [10, 11] with finite element method (FEM) simulations of the electric field around the DBS lead, it has been found that DBS starts to noticeably trigger axons at an electric field magnitude of about 0.2 V/mm at a pulse width of 60 µs and a pulse frequency of 130 Hz [12, 13]. With a combination of FEM and individual axon simulations using a combined cable and Hodgkin-Huxley model, this threshold has been calculated to correspond to activation of axons with an outer diameter including the myelin sheath of 3.4 µm [12]. Very few axons in the human brain are this large however, with outer axon diameters most commonly being around 1 µm even in larger white matter areas such as the medullary pyramid tract [14] or the corpus callosum [15][1]. Smaller axons have been found to require higher stimulation amplitudes both in simulations [12] and experiments [16]. It is thus expected that, while DBS starts to noticeably activate axons at an electric field magnitude around 0.2 V/mm, the effect should be stronger for stronger electric fields closer to the active electrode contacts as smaller and smaller axons become recruited.

The estimation of tissue affected by DBS are obtained from direct combination with axon models or with the simplified model of the electric field magnitude threshold. FEM simulations have been used to e.g. study the impact of perivascular cysts [17], postoperative perielectrode oedema and gliosis [13, 18, 19], electrode design [20], tissue anisotropy [21, 22], and to find optimal target area for Tourette syndrome by the use of voxel-wise comparison of electric field and patient outcome [23]. None of these studies have taken the axon diameter distribution of the brain into account, however.

The aim of this study is to estimate the fraction of activated axons as a function of the electric field magnitude in DBS by comparing the distribution of axon diameters in the human brain from literature [24] with simulations of the electric field magnitude required to activate axons of different diameters.

## Material and methods

### FEM simulation

A FEM simulation in COMSOL Multiphysics 5.5 (COMSOL, Sweden) was used to calculate the external electric potential field, $V_e$ (V), in homogeneous tissue around one active DBS electrode contact of the Medtronic 3389 lead using the equation for steady currents.

$$\nabla \cdot (-\sigma \nabla V_e) = 0 \; (A/m^3) \tag{1}$$

The electric conductivity, $\sigma$ (S/m), in a homogeneous model does not affect the electric potential distribution when using voltage control of the DBS, but was set to 0.1 S/m. The second lowest contact of the lead was set as active with an electric potential of – 1 V relative to ground at the outer border of the tissue geometry. The three remaining inactive contact surfaces, $S$, were set to floating potential

$$\begin{cases} V_e \equiv \text{Constant on } S & \text{(V)} \\ \int \mathbf{n} \cdot \sigma \nabla V_e dS = 0 & \text{(A)} \end{cases} \tag{2}$$

---

[1] Personal communication with Prof. Bente Pakkenberg confirms that the outer axon diameter was measured in this reference.

with **n** being the normal vector to the surface, and the remaining lead surface to electric insulation. The tissue surrounding was modelled as a 105 x 130 x 100 mm³ large cube with approximately 1 700 000 finite elements.

The simulated external electric potential field, $V_e$, and electric field magnitude, $EF = |\nabla V_e|$ (V/mm) was exported along lines corresponding to axons with a tangential horizontal orientation to the middle of the active DBS contact in steps of 0.1 mm from 0.1 mm outside the edge of the contact at a radius, $r = 0.735$ mm to $r = 6.735$ mm measured from the center of the lead (Figure 1).

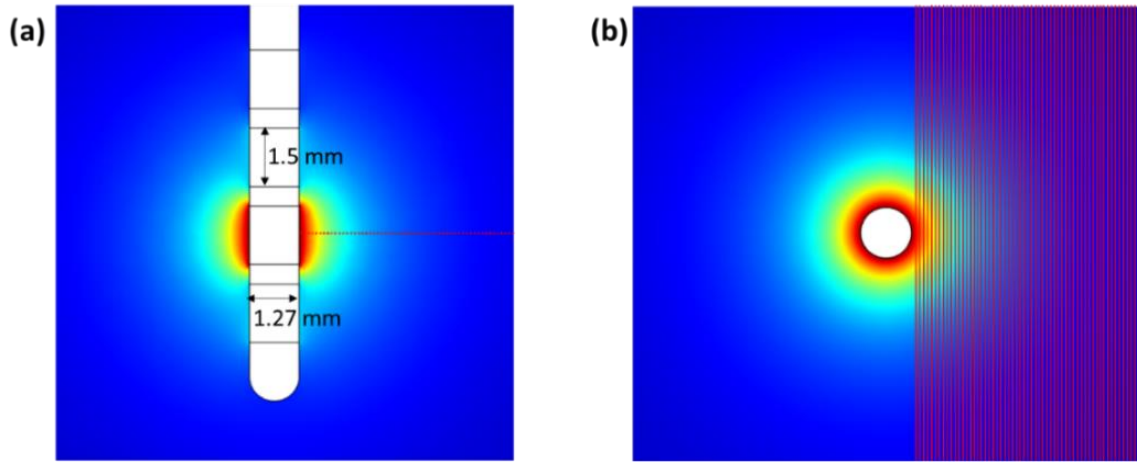

*Figure 1: FEM model and position of exported lines corresponding to the simulated axons in the (**a**) xz and (**b**) xy planes. An electric potential of -1 V was set to the second lowest contact of a Medtronic 3389 lead and the electric potential field (coloured surface field) was exported on lines (red lines) tangential to the contact surface in the xy plane.*

### Axon model

The axon model and corresponding electric circuit is described in Figure 2. Membrane potentials, $V_m$, were calculated in discrete points with every second point in a node of Ranvier and every second being an internode at the middle of each myelin segment using a script in Matlab 2019a (Mathworks, USA) based on original code by Hubert Martens [12].

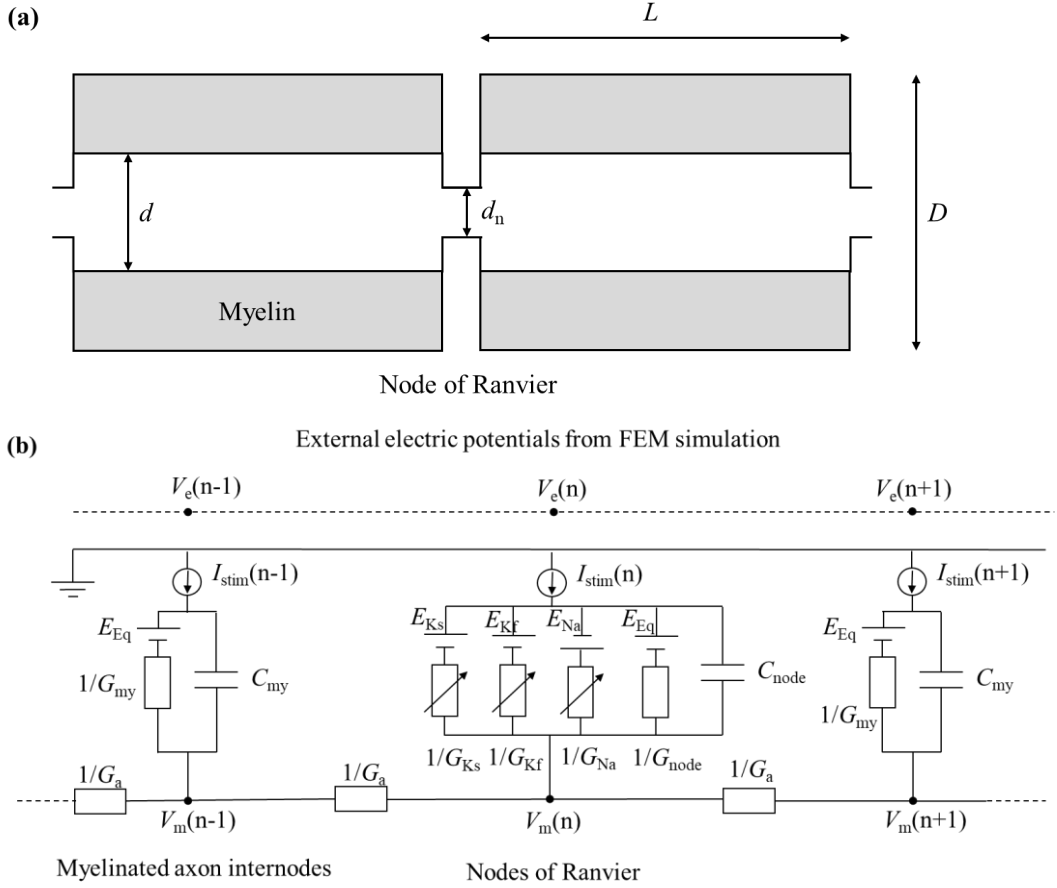

*Figure 2: Axon model. The external electric potential field, $V_e$, from the FEM simulation was used to calculate a stimulation current, $I_{stim}$, for the axon model in each node of Ranvier and myelin internode in order to estimate at what distance from the active DBS contact action potentials are triggered.*

The impact of the external electric potential field, $V_e$ (V), from the DBS lead on the membrane potential in the axons were modelled as a current source, $I_{stim}(n)$ (A), in the discrete points, n, according to

$$I_{stim}(n) = G_A \Delta^2 V_e(n) \text{ (A)} \tag{3}$$

where $G_A$ (S) is the axon conductance of *half* a myelin segment between a node of Ranvier and a myelin internode and $\Delta^2 V_{e,n}$ the second difference of the external electric potential from the FEM simulation. The axon conductance $G_A$ is calculated by:

$$G_A = \frac{\pi d^2}{2L\rho_a} \tag{4}$$

Here, $d$ is the inner axon diameter (µm), $L$ the myelin segment length (mm) between the nodes of Ranvier and $\rho_a$ the resistivity (0.40 Ωm) of the intracellular fluid of the axon [25]. The inner axon diameter was assumed to be related to the outer axon diameter, $D$ (µm), according to [12, 14]

$$d = D \cdot 0.74\left(1 - e^{-D/1.15}\right) \text{ (µm)} \tag{5}$$

and the myelin segment length, $L$ (mm), was assumed to be related to the inner axon diameter according to [12, 26, 27]

$$L = 146 \cdot d^{1.12} \text{ (mm)} \tag{6}$$

The second difference of the external electric potential field is usually called the activation function for axon stimulation and is described by

$$\Delta^2 V_e(n) = V_e(n-1) + V_e(n+1) - 2V_e(n) \text{ (V)} \tag{7}$$

The temporal change in membrane potential in each node, $dV_m(n)/dt$, depends on the sum of the external stimulation current, $I_{stim}(n)$, the current along the axon, $G_A \Delta^2 V_m(n)$ (A), and the currents over the axon membrane into the node (Figure 2b). The membrane potential was calculated according to

$$\begin{cases} C_{node} \frac{dV_m(n)}{dt} = I_{stim}(n) + G_A \Delta^2 V_m(n) + \sum_x G_x(V_x - V_m(n)) \text{ (A)} & \text{Nodes of Ranvier} \\ C_{my} \frac{dV_m(n+1)}{dt} = I_{stim}(n+1) + G_A \Delta^2 V_m(n+1) + G_{my}(V_{Eq} - V_m(n+1)) \text{ (A)} & \text{Myelin sheath} \end{cases} \tag{8}$$

where $C_{node}$ and $C_{my}$ (F) are the capacitance over the axon membrane in the nodes of Ranvier and the myelin sheath internodes, respectively. $G_x$ (S) are the dynamic gated ion conductances and the static leak conductance, $G_{node}$, over the nodes of Ranvier, $V_x$ (V) are the corresponding reversal potentials and $V_{Eq}$ is the equilibrium potential. The gated conductances are the Sodium conductance, $G_{Na}$ (S), and the fast and slow Potassium conductances, $G_{Kf}$ (S) and $G_{Ks}$ (S). $G_{my}$ is the leak conductance over the myelin sheath. The fast potassium current from $G_{Kf}$ has little influence in the action potential of myelinated mammal nerves [28] and may be possible to omit but was still included in the model here.

The conductances $G_x$ are proportional to the node of Ranvier diameter, $d_n$ (µm), which in turn is related to the inner axon diameter, $d$, according to [12, 26, 29]

$$d_n = d \cdot (0.92 - 0.25 \cdot \ln(d)) \text{ (µm)} \tag{9}$$

For details about the membrane conductances and capacitances, including Hodgkin-Huxley gating dynamics, see appendix in Åström et al [12].

Axon simulations were performed for DBS voltages in steps of 0.5 V between 0.5 and 5.5 V, modelled by scaling the $V_e$ solution of the FEM simulation by a factor between 0.5 and 5.5. The outer axon diameters, $D$, were varied in steps of 0.5 µm between 1 and 5 µm, and a pulse width of 60 µs was assumed. Equation (8) was solved in time steps of 20 µs over a pulse cycle of 5 ms, where the external electric potential field, $V_e$, was on for the first 60 µs, using the differential equation solver

ode15s in Matlab (Mathworks, USA) and assuming an initial membrane potential of – 84 mV. Activation of an axon was assumed to occur if the simulated depolarization was sufficient to trigger an action potential, in practice detected by noting a membrane potential $V_\mathrm{m} > 0$ mV at any timepoint during the pulse cycle. Activation distance was noted as the distance from the center of the active DBS contact to the furthest activated axon and a diameter-dependent electric field threshold $EF_\mathrm{t}(D)$ was noted as the scaled electric field magnitude $EF$ from the FEM simulation at that axon in its position closest to the active DBS contact.

### Axon diameter distribution from human brains

An *ex-vivo* study of brain the inner axon diameters, *d*, in three human brains by Liewald et al. [24] was used to calculate an average axon diameter distribution, $p_\mathrm{d}$ (%). A total of 18 axon diameter distributions, with a distribution resolution of 0.1 µm, from the superior longitudinal fascicle, the uncinate/inferior occipitofrontal fascicle, and the corpus callosum were read from the figures of the Liewald study into tabulated form using the free software Engauge Digitizer 4.1.

Corresponding outer axon diameters, *D*, were estimated from the inner axon diameters by least squares fitting (*lsqnonlin*, Matlab, Mathworks, USA) to equation (5), giving an outer axon diameter distribution, $p_\mathrm{D}$ (%). The electric field thresholds for the different diameters were then used to estimate the fraction of activated axons for levels of $EF$ in steps of 0.1 V/mm from 0.1 to 1.8 V/mm. The values of $EF_\mathrm{t}(D)$ were linearly interpolated (*interp1*, Matlab, Mathworks, USA) to obtain the corresponding *D* for each of these $EF_\mathrm{t}$ levels. The fraction, *F* (%), of activated axons was then calculated as the sum of the fractions of all diameters sufficiently large to be activated by each level of $EF$

$$F(EF) = \sum p_\mathrm{D}(EF \geq EF_\mathrm{t}(D)) \; (\%) \tag{10}$$

### Results

The estimated electric field activation thresholds for the different outer axon diameters, *D*, are presented in Figure 3 and were found to increase approximately linearly with $1/D^2$ (Figure 3b). The average axon diameter distributions are presented in Figure 4 and the combined estimate of fraction activated axons, *F*, at different $EF$ are presented in Figure 5. Linear regression (Matlab, Mathworks, USA) between $EF$ and $F$ gave

$$F = 41 \cdot EF - 7.4 \; \% \tag{11}$$

This was thresholded to 0 % at $EF_\mathrm{t0} = 0.18$ V/mm, as electric field magnitudes beneath this value otherwise incorrectly would predict negative activation of axons, and to 100 % at $EF_\mathrm{t100} = 2.62$ V/mm to avoid estimates of more than 100 % activation.

$$\begin{bmatrix} F = 0 \; \% & EF < 0.18 \; \mathrm{V/mm} \\ F = 41 \cdot EF - 7.4 \; \% & 2.62 > EF \geq 0.18 \; \mathrm{V/mm} \\ F = 100 \; \% & EF > 2.62 \; \mathrm{V/mm} \end{bmatrix} \tag{12}$$

An example of the isocontours of *F* for 3 V monopolar stimulation in the FEM simulation is given in Figure 6.

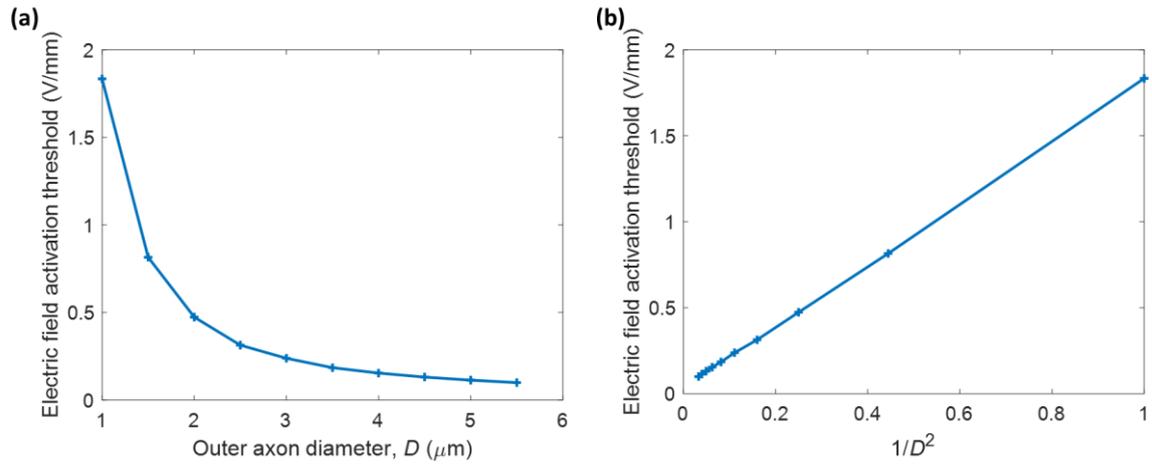

*Figure 3: (**a**) Estimated electric field threshold values, EF$_t$, for different fiber diameters, D, at a pulse width of 60 μs. (**b**) EF$_t$ decreases approximately with $1/D^2$.*

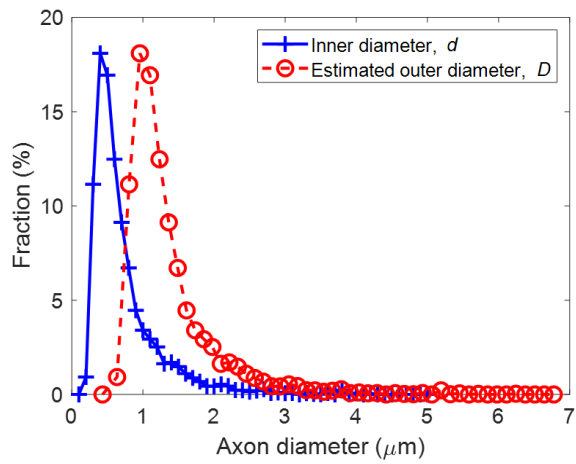

*Figure 4: Average inner axon diameter distribution, p$_d$, in white matter structures in the human brain from Liewald et al [24] and from the inner diameter estimated outer diameter distribution, p$_D$.*

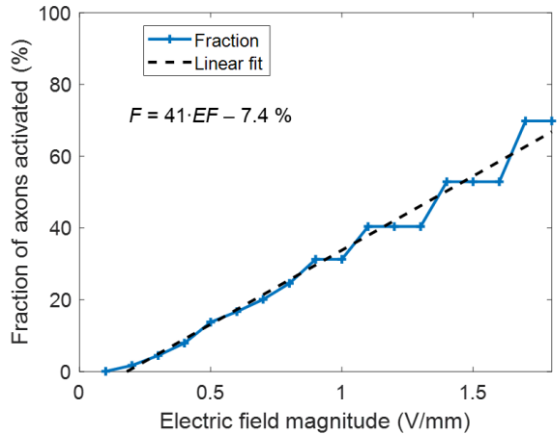

*Figure 5: Estimated fraction of activated axons as a function of electric field magnitude. With a linear regression fit, the fraction of axons activated is estimated to be F = 41·EF – 7.4 % starting at a threshold level of $EF_{t0}$ = 0.18 V/mm.*

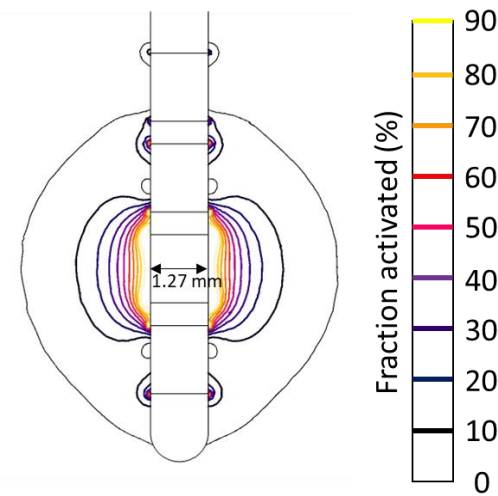

*Figure 6: Estimation of fraction of activated axons from 3 V monopolar stimulation in the FEM simulation. The zero level is assumed to correspond to the 0.18 V/mm EF threshold level.*

## Discussion

In this study, it is estimated that DBS starts to affect the brain tissue at an electric field threshold level of $EF_{t0}$ = 0.18 V/mm with a linear increase in fraction of activated axons as the electric field magnitude increases above $EF_{t0}$. This is in good agreement with comparisons of FEM simulation with experimental studies by Kuncel et al. [10] and Mädler and Coenen [11] where, with adjustments of the Kuncel data to a 60 µs pulse width, $EF_{t0}$ has been estimated to approximately 0.2 V/mm [12, 13]. These comparisons are based on very limited clinical data, however, and more experimental studies for model validation would be desirable. The axon model used in this study has been found to estimate axon activation at lower voltages than the NEURON [30] axon model used by e.g. McIntyre et al [31] with the voltage sufficient for triggering 3.2 µm axons in this study's model only estimating to trigger 5.7 µm and larger axons in the NEURON model [12, 32]. A higher threshold level, $EF_{t0}$, would thus be expected to be estimated if the NEURON model were to be used in this study instead.

The results of this study are particularly relevant to voxel-wise statistical comparisons between simulated DBS and clinical outcomes to calculate improvement maps of optimal target zone and areas responsible for undesired side effects. Most previous studies have assumed a simple all-or-nothing threshold level for the volume of tissue activated (VTA) [23, 33, 34]. This gives an impression of equal effect within the whole VTA when in fact the effect is expected to be stronger close to the active contact where the electric field is stronger and should activate a larger fraction of the axons.

The Liewald study [24] was selected as representative for brain axon diameters as it presents distributions with a high resolution in axon diameter. Riise and Pakkenberg [15] studied the outer axon diameters in the human corpus callosum with coarser distribution intervals and found the most common diameter range to be 0.8-1.1 µm, in good agreement with the from Liewald estimated outer diameters in Figure 4. They had even less large diameter fibers with only about 1-4 % of the axons having a larger diameter than 2.2 µm compared to the estimate of about 10 % in Figure 4. Fiber diameters in the human pyramidal tract of the brainstem have been studied by von Keyserlingk and Schramm [14]. They also found outer diameters of about 1 µm to be most common but found more wide-diameter fibers with about 35 % being larger than 2.2 µm. The brainstem is just above the spinal cord, however. Larger fibers take up more space but allow for faster signal transmission. The brainstem could thus have more large fibers for the purpose of fast long-distance signaling compared to the other parts of the brain.

The used Liewald study [24] contains axon diameters from multiple areas in the brain, but not specifically from common DBS target areas such as the subthalamic nucleus (STN), the thalamus or the globus pallidus interna (GPi). Mathai et al. [35] found a similar inner axon diameter distribution in the GPi and STN in rhesus monkeys as the human inner axon diameter distribution from Liewald, though with somewhat thinner axons in the GPi than the STN. It would be of great interest to obtain similar histology studies for the corresponding axon diameter distributions in humans to investigate if they differ from the ones used in this paper. Explicit measurements of the outer axon diameter would also be of interest to obtain to verify if the assumed relation between inner and outer axon diameter used in this paper (equation 5) is accurate.

The smallest outer axon diameter, $D$, was chosen as 1 µm as this is the approximative starting value for the data [14] underlying the relation between inner and outer axon diameter in equation (5). Smaller axons are expected to require even higher electric field magnitudes than 1.8 V/mm for activation and it is unknown how well they follow the extrapolated values from the regression in equation (12). Such high magnitudes may only be expected very close to the DBS lead contacts at normal DBS settings, however (Figure 6).

The axons were assumed to be oriented parallel to the active electrode surface. Other orientations could be tested and there have for example been studies estimating that orthogonal orientation could induce hyperpolarization instead of depolarization in the axon which would be expected to directly inhibit axon activity [36]. However, as long as the axons are assumed to follow a straight line close to the electrode, they will always be parallel to the contact surface when they are closest to it where the electric field is strongest. They are thus likely to be triggered there even if they experience hyperpolarization further away from the contact.

The simulations do not include the impact of pulse frequency or pulse widths. These effects are probably best studied in clinical experiments on volunteering patients, see e.g. [37], as they readily can be varied and have their impacts studied *in-vivo* without having to make any model assumptions that can introduce unknown errors.

## Conclusions

In conclusion, the fraction of activated axons from DBS in a voxel is estimated to be 41·*EF* – 7.4 % starting at *EF* = 0.18 V/mm. When doing voxel-wise statistics on the effect of DBS, a suitable test would thus be to do linear regression between clinical effect or side effects against *EF* for voxels within the *EF* threshold of 0.18 V/mm.

## Funding
The work was supported by the Swedish Research Council [grant number 2016-03564] and the Swedish Foundation for Strategic Research [grant number BD15-0032].